# Interactive Overlay Maps for US Patent (USPTO) Data Based on International Patent Classifications (IPC)

Loet Leydesdorff,[a]* Duncan Kushnir,[b] and Ismael Rafols [c,d]

**Abstract**
We report on the development of an interface to the US Patent and Trademark Office (USPTO) that allows for the mapping of patent portfolios as overlays to basemaps constructed from citation relations among all patents contained in this database during the period 1976-2011. Both the interface and the data are in the public domain; the freeware programs VOSViewer and/or Pajek can be used for the visualization. These basemaps and overlays can be generated at both the 3-digit and 4-digit levels of the International Patent Classifications (IPC) of the World Intellectual Property Organization (WIPO). The basemaps can provide a stable mental framework for analysts to follow developments over searches for different years, which can be animated. The full flexibility of the advanced search engines of USPTO are available for generating sets of patents and/or patent applications which can thus be visualized and compared. This instrument allows for addressing questions about technological distance, diversity in portfolios, and animating the developments of both technologies and technological capacities of organizations over time.

**Keywords:** map, USPTO, IPC, patent, classification, overlay

[a] Amsterdam School of Communication Research (ASCoR), University of Amsterdam, Kloveniersburgwal 48, 1012 CX Amsterdam, The Netherlands; loet@leydesdorff.net; http://www.leydesdorff.net; * corresponding author.
[b] Environmental Systems Analysis, Chalmers University of Technology, Göteborg, Sweden; duncan.kushnir@chalmers.se.
[c] SPRU (Science and Technology Policy Research), University of Sussex, Freeman Centre, Falmer Brighton, East Sussex BN1 9QE, United Kingdom; i.rafols@sussex.ac.uk.
[d] *Ingenio* (CSIC-UPV), Universitat Politècnica de València, València, Spain.

# 1. Introduction

Alongside scholarly publishing, patents have increasingly become an output of scholarly work. The Bayh-Dole Act of 1980, among other things, granted to universities, small businesses, and non-profit organizations the intellectual property rights to inventions that result from government funding. Other countries followed this legislation and university transfer offices were established to promote knowledge transfer in university-industry relationships. Using non-patent literature references, Narin *et al*. (1997) signaled a more intense and closer linkage between patenting and publishing in several fields of science and technology (Henderson *et al*., 2005). With increased awareness of the emergence of a knowledge-based economy, the patent system became further adapted to the publication system. With the 2011 "American Invents Act," for example, the USA brought its patenting system in line with the rest of the world by changing (as of 2013) from "first to invent" to "first to file" as the basis for granting patents.

More generally, patents can also be considered as indicators of input into the economy. Grilliches (1984) focused on patents as such indicators, and noted the different and sometimes incompatible organization of various statistics (Grilliches, 1994:14). Jaffe & Trajtenberg (2002) then used three million patents and 16 million citations in the USPTO database in a comprehensive study of what these authors called "a window on the knowledge-based economy." However, oatents are indicators of invention; innovation presumes the introduction of inventions into a market. Patents are thus developed in relation to two social contexts: the sciences and markets (Klavans & Boyack, 2008).

Accordingly, patents are classified in terms of technologies and not by products or industries (Jaffe, 1986).[1] The long chain between scientific discovery and technological and/or industrial applications can sometimes be cut short by feedback mechanisms (Kline & Rosenberg, 1986; Von Hippel, 1988), but the two processes are very different in terms of institutional incentives (Rosenberg & Nelson, 1994). The rankings of universities that emerged in the early 2000s, for example, did not include patenting among the indicators of excellence (Leydesdorff & Meyer, 2010).

In summary, patenting is an indicator of industrial activity more than of academic production (Shelton & Leydesdorff, 2012). Whereas scholarly literature is mainly organized into journals, patents are organized into patent classification systems. There are two major classification systems: that used by the US Patent and Trade Office (USPTO), and the International Patent Classifications (IPC) developed by the World Intellectual Property Organization (WIPO) in

---

[1] Schmoch *et al*. (2003) provided a concordance table between International Patent Classifications (IPC) and industrial classifications with the NACE codes ("Nomenclature statistique des activités économiques dans la Communauté européenne") used by the OECD.

Geneva. The latter was first developed for international patenting under the Patent Cooperation Treaty (PCT) that has been signed by most countries of the world since its inception in 1970.

The European Patent Office (EPO) refined the IPC into the European Classification Systems (ECLA) in partnership with the WIPO, the Organization for Economic Cooperation and Development (OECD) in Paris, and the International Patent Documentation Center (INPADOC) in Vienna. The OECD further developed the concept of patent families into "triadic patents," that is, those patents held in common by the USPTO, the EPO, and the Japanese Patent Office (JPO).[2] "Triadic patents" are considered as the highest-quality patents, whereas national patenting can be of mixed quality. Patents can be internationalized under the Patent Cooperation Treaty (of the WIPO) and are then indicated as PCT patents.

The various patenting systems offer firms different routes for patenting: one can patent nationally, internationally (at the WIPO's PCT), or at the regional level such as the EPO. The propensity to patent and the internationalization of patenting can be expected to differ among nations, sectors, disciplines, etc. Among the national patents, USPTO patents are considered the most valuable because of the competitiveness of the US market; the US is the world leader in most technologies (Criscuolo, 2006). As technology indicators, US patents can be assumed to be the most reliable because firms want to secure their intellectual property rights in this largest market.

Like publications, patents can also be distinguished in terms of their numbers of citations, but citation in the case of patenting may mean something different from citation in scholarly literature. In addition to inventor citations, the examiner can attach citations to the front page of the patent in order to ensure coverage of prior art because of the possibility to challenge patents in court (Criscuolo & Verspagen, 2008). Since 2001, the full texts of patents allows one to distinguishes between applicant and examiner citations in US patents, while the latter are asterisked on the front pages of the patents (Alcácer *et al*., 2009).[3] The patenting system is thus regulated more formally than the publication system.

Whereas journals organize scholarly literature in terms of disciplines and subdisciplines as latent variables (which one can bring to the fore using, for example, factor analysis of the citation matrix), patents are categorized manifestly by their patent classification systems. Citations among journals and journal categories have been used successfully to map scientific literature (Klavans & Boyack, 2009). Rafols *et al*. (2010) further developed an overlay technique that enables users to position a document set in terms of its disciplinary affiliations in terms of journal categories. More recently, this system has been refined to the journal level (Leydesdorff & Rafols, 2012).

---

[2] "A patent family is a set of patents taken at various offices to protect a given invention. It is triadic when the invention to which it refers has been the subject of a patent application at the European Patent Office (EPO) and the Japan Patent Office (JPO), and the subject of the issue of a title of ownership at the United States Patent and Trademark Office (USPTO). In other words, a triadic patent protects an invention on the U.S., European and Japanese markets simultaneously." Source: http://www.stat.gouv.qc.ca/savoir/indicateurs/triadiques/index_an.htm.

[3] The program uspto1.exe can be used for downloading the citing patents using the routines available at http://www.leydesdorff.net/indicators/lesson5.htm (Leydesdorff, 2004).

In another context, we have developed a similar overlay system for projecting documents on a basemap of Medical Subject Headings (MeSH) of the Medline database (Leydesdorff, Rotolo, & Rafols, 2012). Using institutional data provided in the bylines of publications and patents, one can furthermore generate overlays to Google Maps for both publications (Leydesdorff & Persson, 2010; Bornmann & Leydesdorff, 2011) and patent data (Leydesdorff & Bornmann, 2012). The organization of a patent map in terms of technological categories such as patent classifications provides a hitherto missing link in this series of studies with the long-term aim of tracing innovations transversally through differently organized domains (Grilliches, 1994; Leydesdorff, Rotolo, & De Nooy, in press; Narin, 2012).

In this study, we turn to mapping the USPTO data in terms of citations among IPC classes. In an earlier attempt, Leydesdorff (2008) explored mapping WIPO data in terms of IPC co-classifications, but noted that the hierarchical structure introduced by the thesaurus made it difficult to map the co-classification structure at the aggregated level. Indexer effects are generated, for example, when classes are split (or otherwise changed) because they grow too large. Such effects can have an uncontrolled impact on co-classifications.

Classifications make discrete cuts, whereas the network of citation relations can vary in density within and across clusters (Kay *et al*., 2012). In other words, the citation network among the IPC classes is heterarchical, whereas the IPC provides a hierarchical representation (with one less degree of freedom). Because of the additional degree of freedom in networks when compared with hierarchies, one can expect that the citation network is less sensitive to misclassifications than the co-classification network (Rafols *et al*., 2010: 1887).

Co-classifications have more often been used for measuring the "technological distances" between patenting units such as firms or nations (e.g., Breschi *et al*., 2003; Dolfsma & Leydesdorff, 2011; Jaffe, 1986, 1989). More recently, several academic groups have proposed organizing patent data in terms of aggregated citation structures among IPC classes (Kay *et al*., 2012; Schoen, 2011). Schoen *et al*. (2012) use a database derived from the comprehensive PatStat database, but selected by the Corporate Invention Board (CIB) and used by the Institute of Prospective Technology Studies of the European Commission. Furthermore, these authors construct a classification of their own.

Kay *et al*. (2012) used EPO data from 2000-2006, but varied the number of digits in the IPC hierarchy in order to optimize the sizes of the categories for the sake of mapping. In our opinion, one can normalize for size differences among distributions using, for example, the cosine. Our approach is closest to that of Boyack and Klavans (2008:181), who developed a USPTO patent map based on co-classifications of 290 IPC classes.

USPTO data were collected by one of us for the years 1976-2011 (approximately 4.2 million patents), and the approximately 39 million citation relations were organized in terms of IPC classes. The aggregated citation matrices at the 3-digit and 4-digit level of IPC are normalized using the cosine as a similarity measure among citation distributions in different classes (Ahlgren *et al.*, 2003). Jaffe (1986, 1989:88) defined "technological proximity" and "technological distances" in terms of the cosine measure that we also use standardly in other maps. Our purpose is to make available an interactive basemap comparable to the previously constructed basemaps for journals, journal categories, and MeSH categories, and to leave the user as much flexibility as possible for further adjustments. All necessary programs and data needed are in the public domain, including the programs VOSViewer and Pajek that are used for the visualization (De Nooy *et al.*, 2011; Van Eck & Nees, 2010).

USPTO data has been available online in html format since 1976.[4] At the time of this study, 2011 was the last complete year. All valid citations from the period 1976-2011 among IPC classes at the 3-digit and 4-digit levels are used for the basemaps; two basemaps are accordingly provided. A routine is available online (at http://www.leydesdorff.net/ipcmaps) to assist the user in downloading specific sets from the USPTO database[5] and organizing each set as an overlay to the basemaps. As with our previous maps, we use the "citing" side of the cited/citing matrix among the classes for the analysis, because in the "cited" direction older patents may be prevalent, whereas the analyst is chiefly interested in the current state of a unit of analysis (e.g., a country or a technology) publishing and citing patents in its knowledge base. In a later state, we envisage extending the system to include the "being cited" counts, as was done—using the top-quartiles with different colors—in case of the overlays of patent statistics to Google Maps (Leydesdorff & Bornmann, 2012).

## 2. Methods and materials

The complete set of USPTO patents for the period 1976-2011 was downloaded from Google on February 12, 2012. This set contains 4,597,127 patents ranging from 75,544 patents granted in 1976 to 247,727 in 2011 (Figure 1). These are the so-called technical patents; design patents and genetic sequences were excluded. References in technical patents to design patents or genetic sequences are also not included in this data.

---

[4] EPO data are available online, but there are limitations to the searches (above 500 documents) and some of this data is in the pdf format. PCT patents of WIPO are available online (Leydesdorff, 2008), but as noted of lower technological and market quality than USPTO patents (Shelton & Leydesdorff, 2012).
[5] The two databases can be found at http://patft.uspto.gov/netahtml/PTO/search-adv.htm for granted patents, and http://appft1.uspto.gov/netahtml/PTO/search-adv.html for patent applications, respectively.

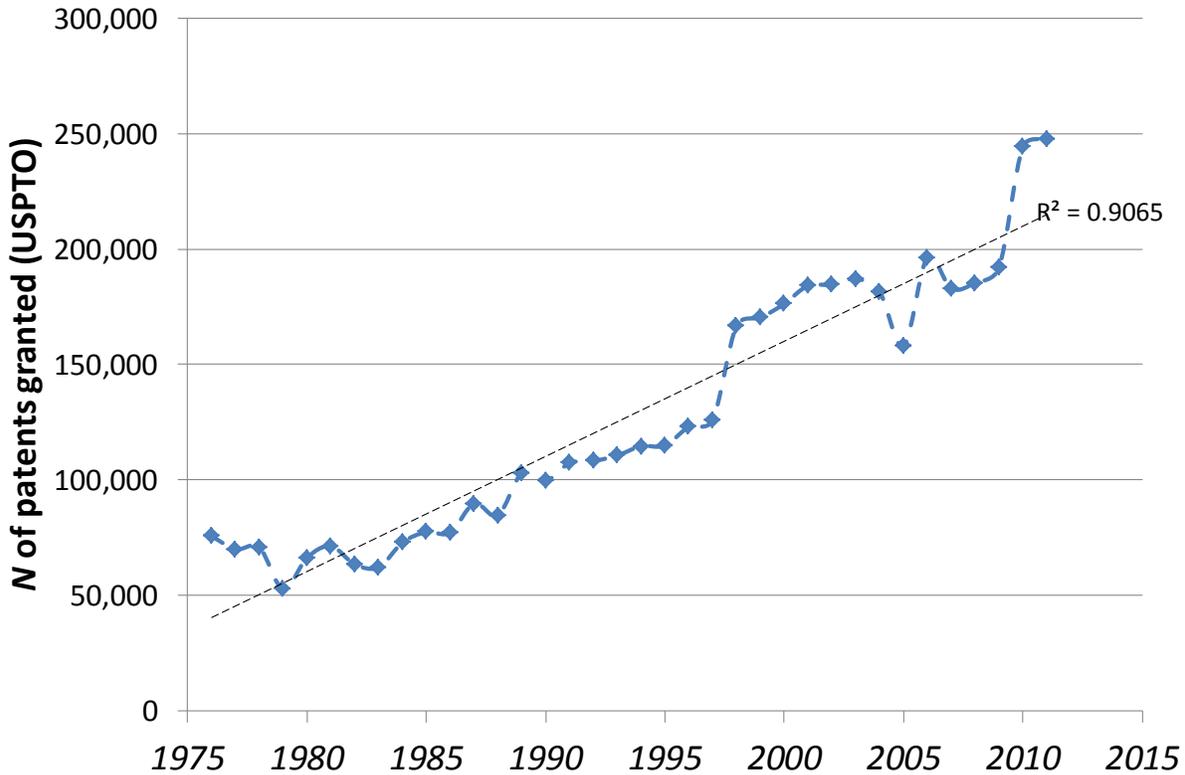

**Figure 1**. *N* of technical patents (granted) in the USPTO database, 1976-2011.

The number of IPC-classes per patent can range from one to more than twenty in exceptional cases. However, only the primary classes of patents were used for generating the basemaps. Among these patent classes, citations were counted at both 3-digit and 4-digit levels and organized in an (asymmetrical) citation matrix (that is, "citing" versus "cited"). Table 1 provides the descriptive statistics.

|  | *3 digits* | | *4 digits* | |
| --- | --- | --- | --- | --- |
|  | USPTO data (including database mistakes) | After correction | USPTO data (including database mistakes) | After correction |
| *IPC Classes* | 817 | 124 (129 in IPC) | 4,126 | 630 (637 in IPC) |
| *N of links* | 28,599 | 13,541 (47.4%) | 253,049 | 176,972 (72.8%) |
| *N of citations* | 39,278,933 | 39,124,366 (99.6%) | 39,286,577 | 38,824,390 (98.8%) |

**Table 1**. Descriptive statistics of IPC classes and their citation relations in the USPTO data (1976-2011) at three- and four-digit levels.

The total number of citations in the matrix is on the order of 39 million (Table 1), independently of whether this is measured at the 3-digit or 4-digit level. Note that the correction for mistakes in the database has a considerable effect (52.6% and 27.2%, respectively) on the number of links, but not on the number of citations. The error is negligible at the 3-digit level (0.4%) and still relatively small (1.2%) at the 4-digit level. As noted above, errors in the classifications can often be considered as disturbances in terms of the bibliographic links at the network level. Furthermore, the IPC contains several classes such as "C99" (at the 3-digit level) which are labeled as "subject matter not otherwise provided for in this section," but these classes are not used by USPTO. Thus, we use 124 of the 129 available IPC at the 3-digit level, and 630 or the 637 at the 4-digit level. In sum, we analyze the data listed in the right columns of 3-digit and 4-digit data in Table 1, after deletion of the misclassifications.

The two citation matrices (for the 3- and 4-digit levels; available from the website) are input into SPSS v.20, so that cosine-normalized matrices can then be computed and exported. The cosine is a non-parametric similarity measure; it has the advantage of being insensitive to the large numbers of zeros in the vectors of sparse matrices. The cosine can also be considered as a non-parametrized Pearson correlation coefficient. Jaffe (1989:89; cf. Jaffe, 1986:986) proposed using the cosine to measure technological distances, but he used co-classifications. As noted, cosine values were computed from the citing side of the matrices because "citing" is the active variable, whereas in the "cited" dimension the complete archive can be represented.

In previous mappings, we generated basemaps from cosine matrices using the spring-embedded algorithm of Kamada & Kawai (1989) in Pajek. Since cosine values tend to be larger than zero in (almost) all cells of the matrix, a threshold had to be set in order to make a grouping of the categories visible; otherwise, almost all vertices are connected by a single (largest) component. Using the MDS-like solution of the program VOSViewer, however, a threshold is not needed, since the algorithm uses all quantitative information for the mapping (Van Nees *et al*., 2010). Leydesdorff & Rafols (2012) noted that further normalization within VOSViewer did not disturb the maps on the basis of cosine-normalized matrices.

Because of the additionally available network statistics in Pajek (and similar programs for network analysis), we shall extend the options by providing base maps and overlay files in the Pajek-format in a later section. (Section 4: Alternatives.) Other programs such as UCINet and Gephi can read Pajek files. VOSViewer and Gephi have solved the problem of the cluttering of labels in the visualization (Leydesdorff *et al*., 2011); Pajek and VOSViewer are most convenient for making overlay maps, but differ in terms of the possible visualizations and clustering algorithms available (Leydesdorff & Rafols, 2012). For the sole purpose of visualization, VOSViewer is probably the first option.

VOSViewer provides its own clustering algorithm based on modularity optimization (Blondel *et al*., 2008; Newman, 2004; Waltman *et al*., 2010). We use this algorithm as the default in VOSViewer. Both in the 3- and 4-digit case, the results are conveniently in five main clusters and therefore colors; all IPC classes (124 and 630, respectively) are included in both cases. The user is able to replace the clustering and colors (see Leydesdorff & Rafols [2012] for more detailed instruction). For example, one may wish to color the maps according to the eight top-level categories of IPC (A to H). The text files that contain the mapping information (available at the website) can be edited. VOSViewer (at http://vosviewer.com) can be either web-started or downloaded from the internet and used locally.

The labels at the 4-digit level are sometimes long, and this may affect the readability of the maps. VOSViewer uses 30 characters as a default, but this can be adapted within the program interactively. One can also change the color selection interactively or turn off the prevention of blurring of the labels that is provided as a default. In the case of long labels, we follow the common practice of using the IPC (sub)headings by cutting off at a maximum of 75 characters, with three dots after the right-most space in the string. The user may edit the files differently if so wished. A default of 30 characters works without problems in the 3-digit case, but may require some editing in the 4-digit map. Note that the length of the strings can affect the visibility of individual labels because the program optimizes readability.

The maps provide a representation of distances between categories and the diversity among categories in each specifically downloaded set of patents. In addition to the distance on the visible map, $\|x_i - x_j\|$, the (technological) distance in the data between each two classes can also be denoted analytically as $d_{ij} = (1 - cosine_{ij})$. The mapping program projects the multivariate space—spanned by the citing patents as vectors of the matrix—into the two dimensions of the map by minimizing a function such as Kruskall's (1964) stress.[6]

The technological distances between categories ($d_{ij}$) can also be summed and normalized in accordance with the weight of the respective categories using their proportions in the distribution. This leads us to the Rao-Stirling measure of diversity ($\Delta$) that is sometimes also called "quadratic entropy" (e.g., Izsáki & Papp, 1995; cf. Rao, 1982):

$$\Delta = \sum_{ij} p_i p_j d_{ij} \qquad (1)$$

---

[6] Kruskall's formula (1964) is expressed as follows:

$$S = \sqrt{\frac{\sum_{i \neq j}(\|x_i - x_j\| - d_{ij})^2}{\sum_{i \neq j} d_{ij}^2}}$$

In addition to summing the differences between the visually available distances and the algorithmic distances, this stress value is normalized at the level of the system of distances. Regrettably, VOSViewer does hitherto not provide a stress value for the (deterministic) visualization.

Stirling (2007:712) proposed this measure as a summary statistics for analyzing diversity and interdisciplinarity in studies of science, technology, and society (Leydesdorff & Rafols, 2011a; Rafols and Meyer, 2010). The routine writes the values of this parameter locally to a text-file ("ipc_rao.txt") at both the 3-digit and 4-digit level.

In summary, we operationalize the technological distance between patent classes as the complement of the cosine between the distributions of citations, and are then able to compute diversity in terms of Equation 1. The cosine matrices for both 3-level and 4-level IPC are available at the webpage, and if these files are present, the routine writes in each run a file "ipc_rao.txt" that contains the diversity measures given the two matrices at the respective digit-levels.

Note that there is no hierarchical relation between the two maps because a patent may contain a single category at the 3-digit level, under which many different 4-digit categories can be subsumed that co-vary with other 4-digit categories at this level. In the case of the overlays, IPC classes will be used proportionally, that is, as a proportion of the total number of IPC classes attributed to each patent, so that each patent contributes a sum value of unity to the overlays. Both the numbers of patents and the fractional counts are made available in a file "vos.dbf" at the occasion of each run. Note that this file is overwritten by a next run!

## 3. Results

*3.1. basemaps at the three- and four-digit level*

The generation of the basemaps from the cosine matrices is straightforward. Figure 2 shows the basemap of 124 IPC classes at the 3-digit level, and Figure 3 shows 630 classes at the 4-digit level. In the latter case we used the heat or density map as an alternative representation.

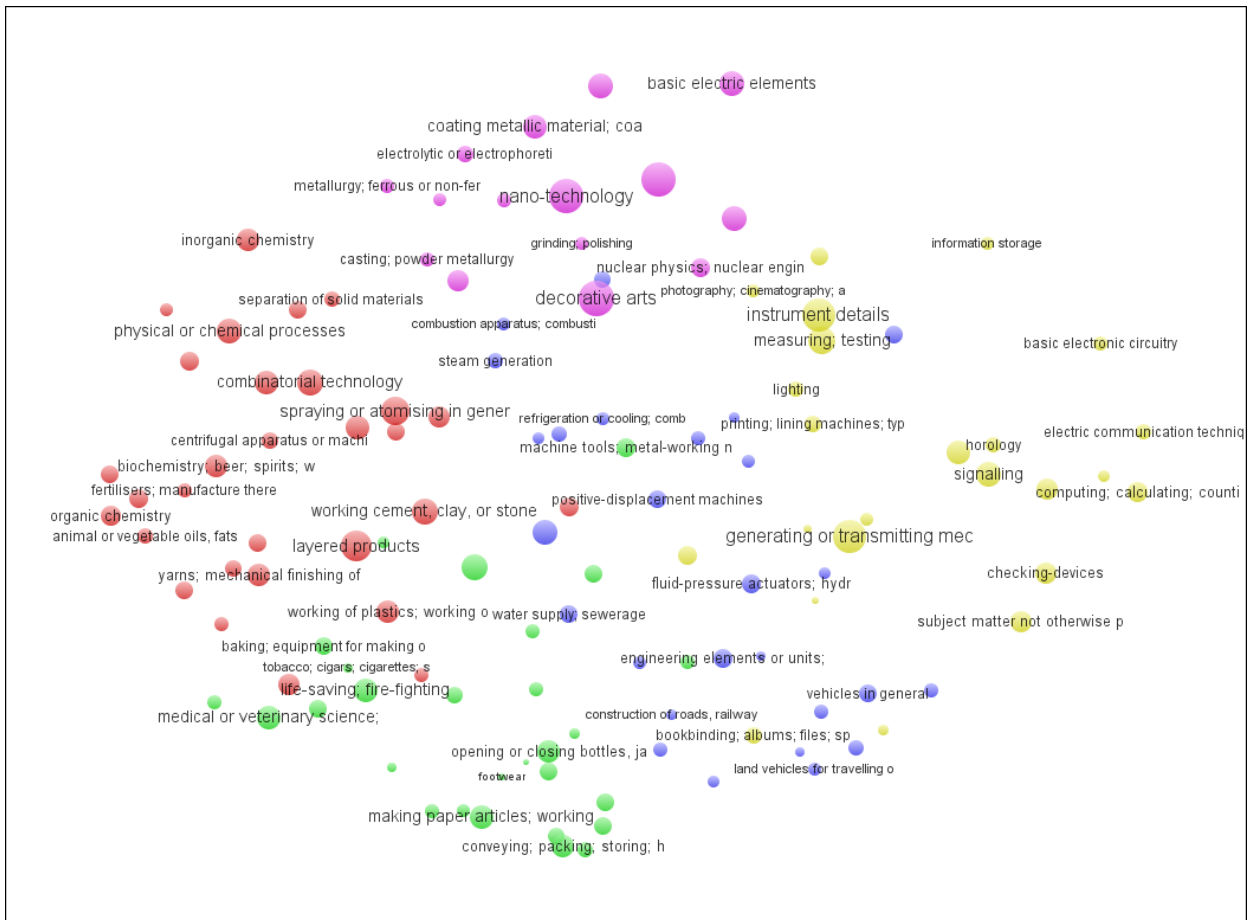

**Figure 2**: Basemap of 124 IPC categories at the 3-digit level, using approximately 39 million citations among 4.2M US patents (1976-2011); cosine-normalized "citing"; VOSViewer used for the visualization. This map can be accessed interactively at http://www.vosviewer.com/vosviewer.php?map=http://www.leydesdorff.net/ipcmaps/ipc3.txt .

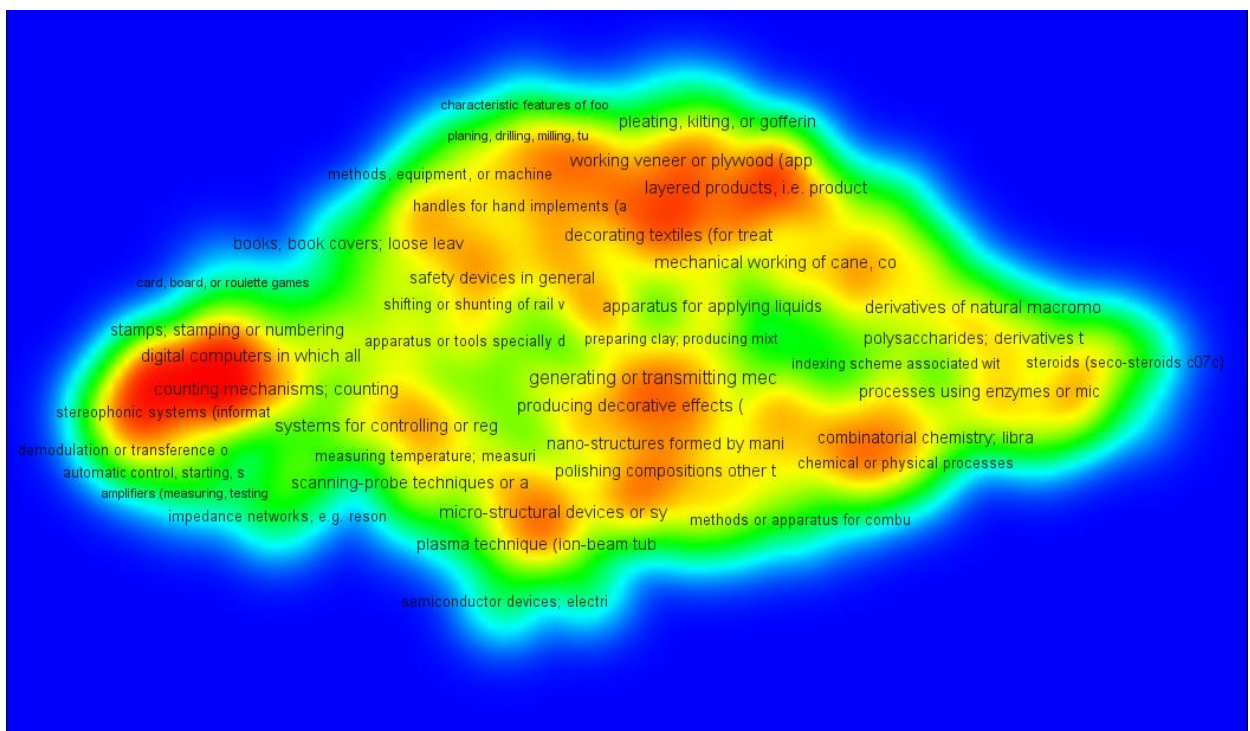

**Figure 3**: Density map for normalized citation relations among 630 IPC categories at the 4-digit level (30 characters used); VOSViewer used for the visualization. This map can be accessed interactively at
http://www.vosviewer.com/vosviewer.php?map=http://www.leydesdorff.net/ipcmaps/ipc4.txt

| Cluster | IPC 3-digits | | IPC 4-digits | |
|---|---|---|---|---|
| | N | Designation | N | Designation |
| 1 (red) | 30 | Chemistry | 215 | Machines, vehicles |
| 2 (green) | 28 | Tools; machinery | 145 | Chemistry |
| 3 (blue) | 27 | Engineering | 104 | Manufacturing |
| 4 (yellow) | 24 | Instruments | 94 | Instruments |
| 5 (pink) | 15 | Materials | 72 | Materials |
| | 124 | | 630 | |

**Table 2:** Five clusters using 3 digits and 4 digits of the IPC, respectively; clustering generated by VOSViewer.

Table 2 provides cluster designations based on the most frequently occurring words in the labels of the IPC at the respective level. The 3-digit labels seem sufficiently precise for high-level analysis (e.g. policy analysis of countries), whereas 4-digit labels may be needed for specific technology analysis. As might be expected, the resolution among the labels at the 3-digit level is higher than at the 4-digit level. Note that the 3-digit and 4-digit maps exhibit the same structural dimensions but flipped both horizontally and vertically; each solution may freely be rotated and translated in the MDS domain.

*3.2 Overlay maps*

One can generate the input for overlays by using, for example, the interface of USPTO Advanced Searching for (granted!) patents at http://patft.uspto.gov/netahtml/PTO/search-adv.htm . First, the search must be formulated so that a recall of more than fifty patents is generated. By clicking on "Next 50" one finds the information used by the dedicated programs ipc.exe and uspto1.exe (available at http://www.leydesdorff.net/ipcmaps) to automate further searches, and to organize input files for VOSViewer and Pajek thereafter. Any search string compatible with the USPTO database will do the job. We provide examples below.

In patents with a sequence number in the retrieval higher than 50, one copies (Control-C) the complete string in the browser after opening this patent (for an example and instruction see also at http://www.leydesdorff.net/ipcmaps). All information about the search is embedded in this string and can be used by the program uspto1.exe to download the set. The routine uspto1.exe is optionally called by ipc.exe.

The patents are downloaded and further organized in the same folder of the disk, and two output files (named "vos3.txt" and "vos4.txt," respectively) can be used directly by VOSViewer as input for generating overlays. In addition to "ipc.exe", the user should first download also the files "ipc.dbf" and "uspto1.exe" from the same website into the same folder because these files are also required. If the files "cos_ipc3.dbf" and "cos_ipc4.dbf" are also downloaded from the website, the routine writes a local file "ipc_rao.txt" containing the Rao-Stirling diversity values for three and four digits, respectively.

The program ipc.exe opens with the option to download patents in this run or use patents downloaded in a previous run. Note that patents from a previous run are overwritten in the case of a new download; one should save them elsewhere for future use. The USPTO limits the number of downloads to one thousand, but one can begin subsequent downloads at 1001, 2001, etc. The program accepts subsequent numberings. In this case, one should use uspto1.exe directly, because ipc.exe overwrites results from previous runs.

In the output of the routine ("vos3.txt" and "vos4.txt") the IPC classes assigned by the USPTO to patents in the download are organized into the map files of VOSViewer on the basis of counting each IPC attribution in proportion to the total number of classes attributed to the patent at the level in question. In the "label view" of VOSViewer, the empty classes are made visible as little grey dots for the orientation of the user, but these classes are not labeled. The classes in use are normalized proportionally to the logarithm of the number +1. The "+1" prevents single occurrences or fractions smaller than unity from disappearing.[1]

Figure 4 shows the density map of all patents granted in 2007 with a Dutch address for the inventor. This same data was used for an overlay to Google Maps by Leydesdorff & Bornmann (2012, at pp. 1446 ff.). One can edit the input files or use the interactive facilities of VOSViewer to further enhance the representation.

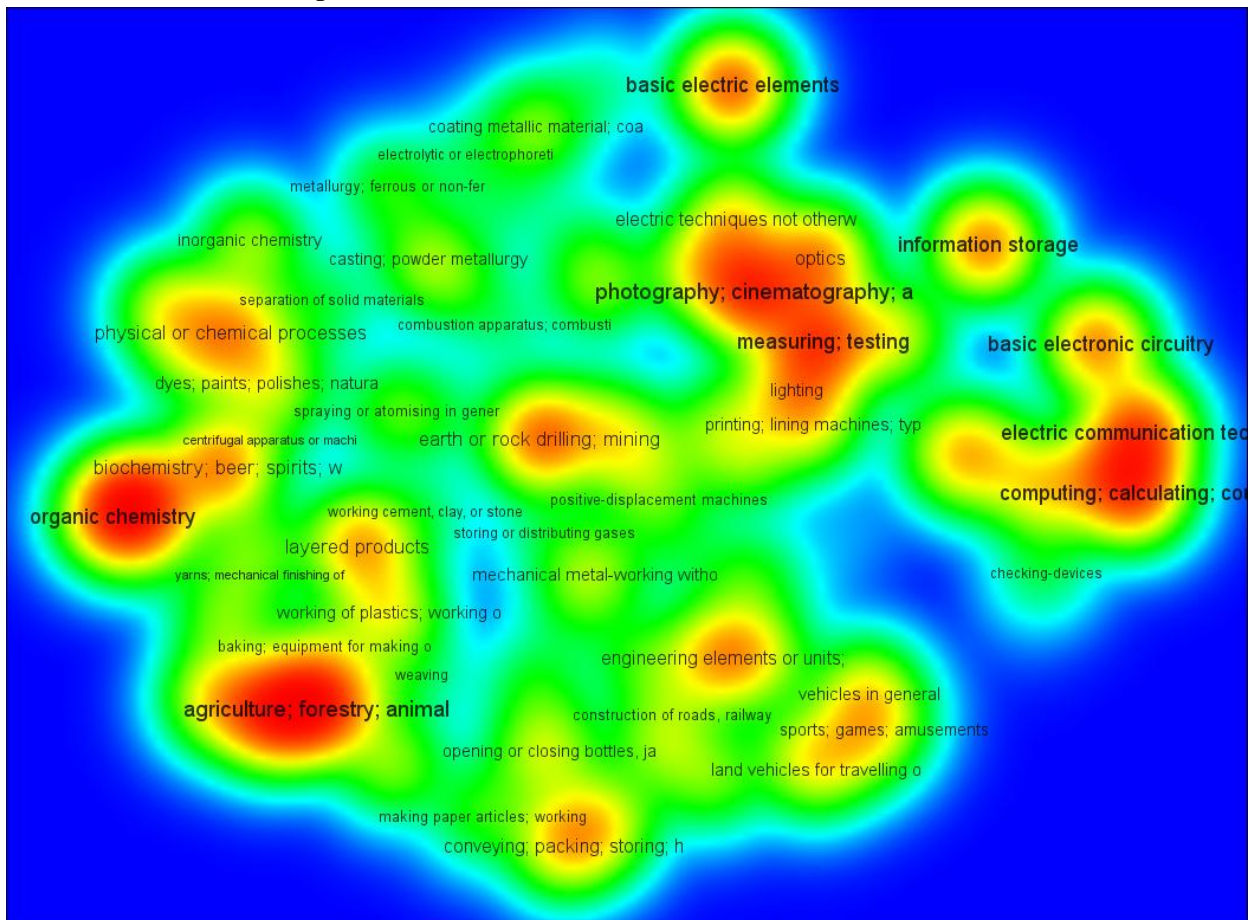

**Figure 4**: Heat map of 1,908 US patents with an inventor in the Netherlands, and publication date in 2007 at the 3-digit level; 3018 IPC classes (cf. Leydesdorff & Bornmann, 2012, pp. 1446 ff.; Rao-Stirling diversity Δ = 0.869. This map can be accessed interactively at http://www.vosviewer.com/vosviewer.php?map=http://www.leydesdorff.net/ipcmaps/vos3.txt&view=2

VOSViewer shows the labels of the most prominent classes in the sets under study. (One can also turn this off.) As noted, the user can modify the clustering classification; our labels correspond to the labels in the (most recent) IPC 2012 version. The quantitative information about the results of each run are stored both at the 3-digit and 4-digit levels in a file "vos.dbf" for further (e.g., statistical) analysis.

---

[1] This addition of unity is needed because log(1) is zero. Fractional counts may be smaller than unity.

The USPTO interface offers a host of possibilities for combining search terms such as inventor names and addresses, applicants, titles, abstract words, issue dates, etc., that can be combined with AND and OR operators. Furthermore, a similar interface is available at http://appft1.uspto.gov/netahtml/PTO/search-adv.html for searching patent *applications*. For example, 3,898 applications with a Dutch inventor address can be retrieved by using "icn/nl and apd/2007$$" as a search string in this database. (Unlike "isd" for "issue data" above, "apd" is the field tag for "application date".)

A second routine named "appl_ipc.exe" operates similarly to "ipc.exe", but parses downloads from this interface of patent applications. (The format is somewhat different.) As above, any webpage-address can be harvested at the 51$^{st}$ patent application, for example, and used for the download (which is strictly similar to the one using ipc.exe). Figure 5 shows a map of the applications corresponding to Figure 4 in terms of search domains, but this time the label view is used for the display. One analytical advantage of using applications is that applications follow the research front, whereas the process of examining and granting a patent can take several years.

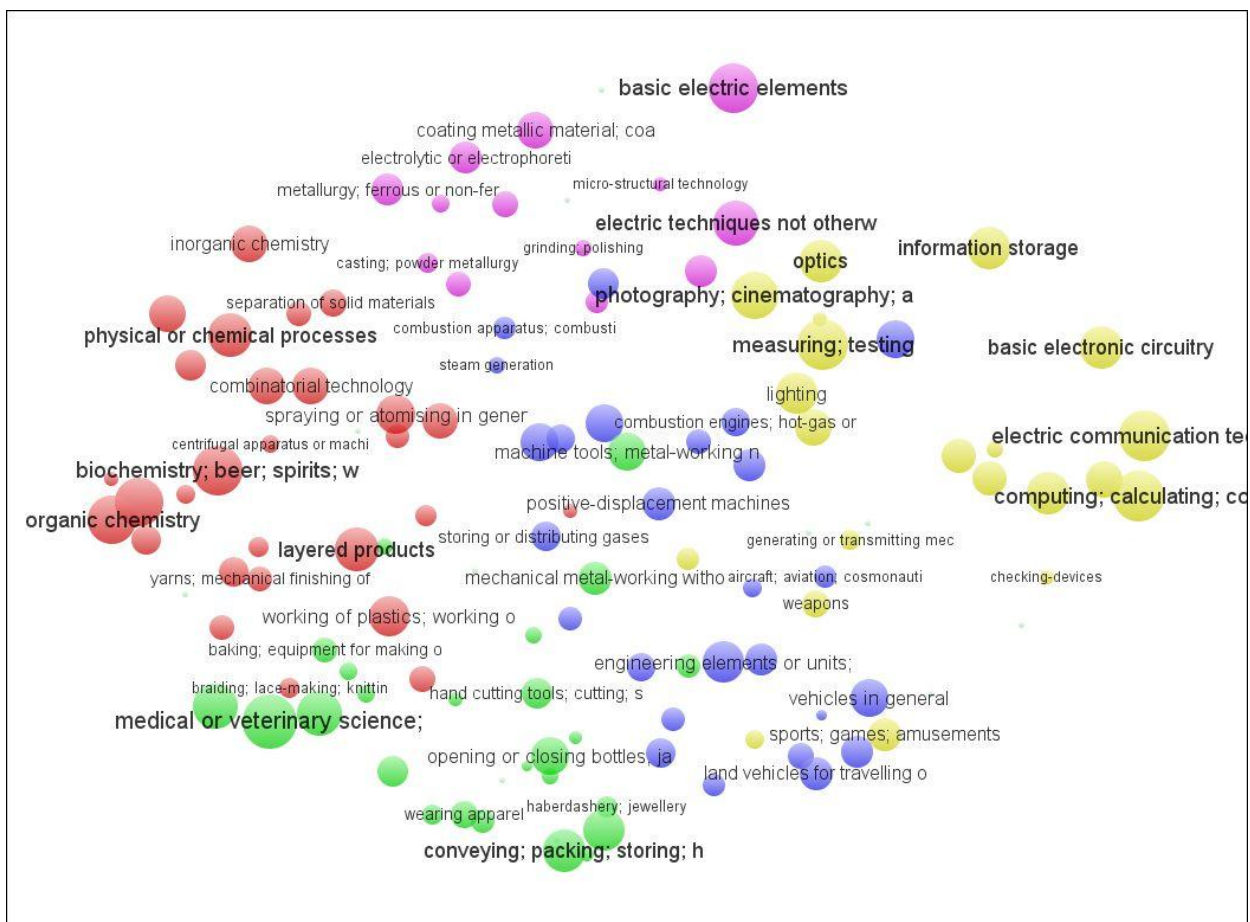

**Figure 5**: Overlay map of 3,898 US patent *applications* with at least one inventor address in the Netherlands, and application date of 2007 at the 3-digit level; 7,872 IPC classes; Rao-Stirling diversity Δ = 0.848.
http://www.vosviewer.com/vosviewer.php?map=http://www.leydesdorff.net/ipcmaps/vos3appl.txt

The distribution of patents granted and patent applications in 2007 with a Dutch address (Figures 4 and 5, respectively) correlate significantly ($r = .753$; $p < .01$), and the Rao-Stirling diversity is almost similar.[2] Since the basemaps remain by definition the same, it may sometimes be difficult to detect these differences upon visual inspection. However, one can animate developments over different years or among different domains because the basemap stabilizes the framework.

By mapping the same searches for different years, for example, and copying the results into a program such as PowerPoint, one can animate the visualizations. The positions will be stable, but the size of the nodes and the most prominent labels will change. We use the option "normalized weights" as default in vos3.txt and vos4.txt so that the user can compare results across years and otherwise. However, for esthetic optimization by VOSViewer, one may wish to turn this option into "weight" in a single-case study. As noted, the colors and sizes of nodes that one wishes to highlight can be adapted by editing these lines in the input files to VOSViewer. Thus, new developments can be made visible in animations. For example, one can add exclamation marks to the labels.

---

[2] The corresponding value at the 4-digit level is: $r = .648$ ($p < .01$).

## 4. Alternatives

Table 3 provides an overview of the various teams that provide or have provided alternative maps of patents. None of the alternative programs offer the interactivity that is offered by our programs. As noted, the USPTO offers the most flexible and public interface, and there are also theoretical reasons for using USPTO data.

|  | Kay *et al.* (2012) | Schoen *et al.* (2012) | Boyack & Klavans (2008) | This study |
|---|---|---|---|---|
| Database | EPO 2000-2006 | CIB 1986-2006 | USPTO 2002-2006 | USPTO 1976-2010 |
| Nr of patents | 760,000 | > 6 million | 907,500 | 4,597,127 |
| Nr of IPC classes | 486 IPC7 (modified) | 389 | 290 | 124 3-digit 630 4-digit |
| Nr of citations | 28,457,418 | 0 | 0 | appr. 39M (see Table 1) |
| Similarity measure | Cosine | Non-normalized | K50 measure | cosine |
| Visualization | Pajek | Gephi | VxOrd | VOSViewer, Pajek |

**Table 3**: Comparison among alternative programs for mapping patents.

Leydesdorff (2008) also discussed mapping one year of PCT-patents at WIPO using co-classifications. Boyack & Klavans (2008) used co-classifications, but the objective of these authors was to analyze science-technology interrelations in terms of inventors and scholarly authors. Their map (at p. 181) also explores relations to industrial sectors (cf. Hinze *et al.*, 1997; Schmoch *et al.*, 2003).

Schoen *et al.* (2012) uses a selection of patents based on firm-selection within and outside the European Union. These authors also decided for co-classification after comparing the results with citation analysis. Kay *et al.* (2012) argue (with us) for using citation analysis among patent categories and against using co-classifications, but these authors unfortunately used the previous version of IPC (v. 7) and manipulated the categories so that they are equal sized. In our opinion, the different sizes can be normalized using the cosine.

The programs differ also in terms of the visualization techniques. In our opinion, Gephi and VOSViewer offer superior visualization techniques (Leydesdorff *et al.*, 2011), but Gephi and Pajek/UCINet offer network statistics. However, the comparison made us realize that with little effort we could also make our outputs compatible with Pajek, and via Pajek also for Gephi (which read Pajek files). This offers additional flexibilities such as using algorithms for community detection among a host of other network statistics which are available in Pajek and Gephi, but not in VOSViewer.

Baseline maps for Pajek in the 3-digit and 4-digit format can be retrieved at http://www.leydesdorff.net/ipcmaps/ipc3.paj and http://www.leydesdorff.net/ipcmaps/ipc4.paj, respectively. The available baselines were in this case partitioned—for didactic purposes—using Blondel *et al.*'s (2008) algorithm for community finding. Without a threshold for the cosine, the visualization is not informative. Using a threshold of cosine > 0.2, the largest components contain 109 of the 124 classes at the 3-digit level (11 communities; $Q = 0.529$), and 605 of the 630 classes at the 4-digit level (21 communities; $Q = 0.681$). However, the experienced user should change these base maps and their partitioning or coloring using the options available in Pajek (or Gephi).[3]

Both routines (ipc.exe and appl_ipc.exe) provide two vector files (ipc3.vec and ipc4.vec, respectively) with the same values—"weights" in VOSViewer—as used above; that is, the fractionally counted patents/IPC class. These vector files can be used for the visualization of the patent contributions to IPC classes in the downloaded sample(s) under study.

Additionally, the so-called cluster files ipc3.cls and ipc4.cls (in the Pajek format) enable the user to label the sample under study exclusively in the Draw screen under Options>Mark Vertices Using>Mark Cluster only. As an example, Figure 6 provides a visualization of the same set as used in Figure 4 with VOSViewer, but now using the spring-embedded algorithm of Kamada & Kawai (1989) and exclusive labeling for one of the communities of IPC classes in this set.

---

[3] The full cosine matrices are available at http://www.leydesdorff.net/ipcmaps/cos_ipc3.dbf and http://www.leydesdorff.net/ipcmaps/cos_ipc4.dbf for those users who wish to be able to work with cosine values lower than the current threshold of 0.2.

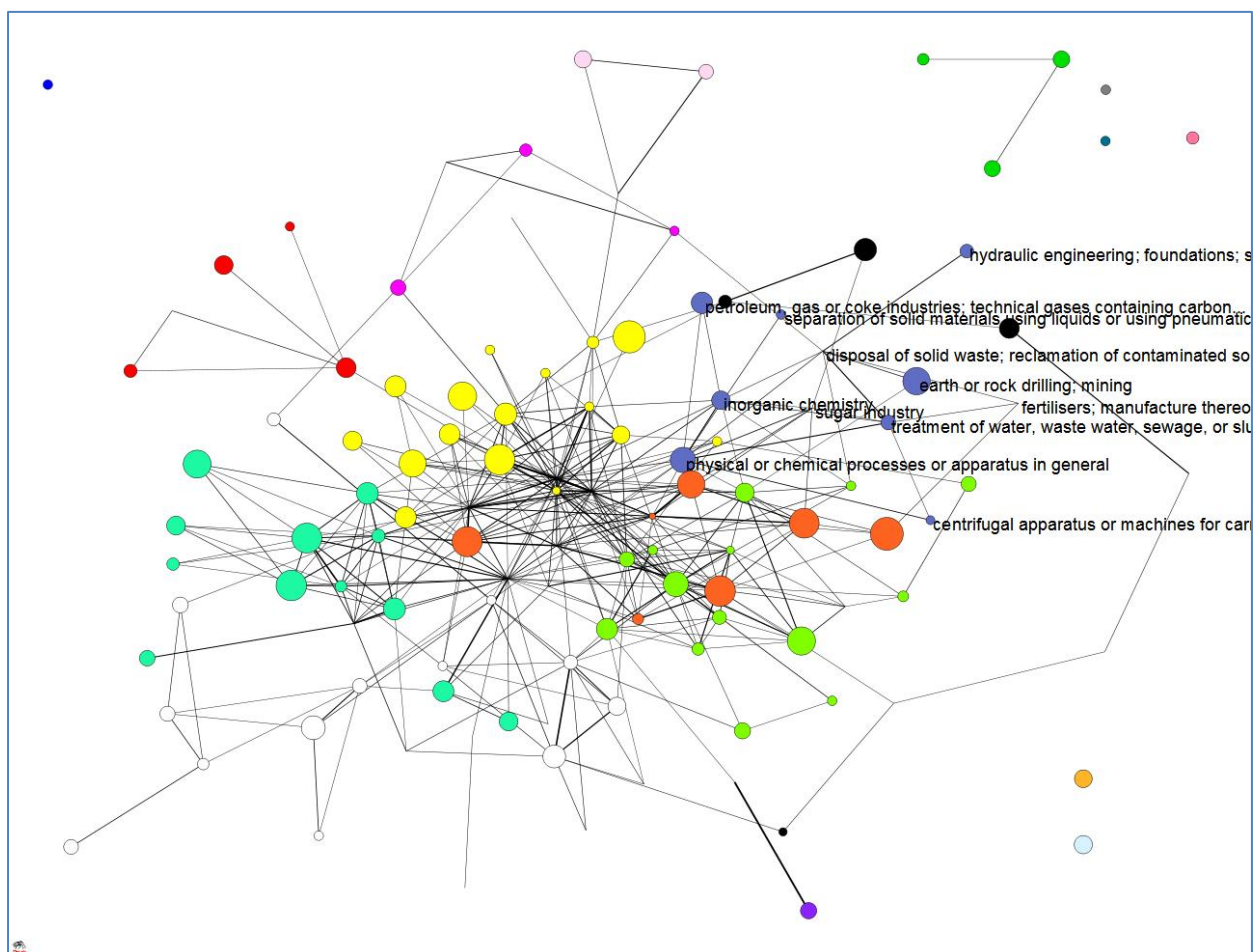

**Figure 6**: Equivalent map of Figure 4, but using Pajek: 1,908 US patents with an inventor in the Netherlands, and publication date in 2007 at the 3-digit level; 3018 IPC classes. Layout according to Kamada & Kawai (1989); coloring according to Blondel *et al*. (2008); one community (nr. 9) is labeled.

All these programs enable the user to export the visualizations as "scalable vector graphics" (SVG). SVG-files can be edited in considerable detail using the freeware program InkScape (at http://inkscape.org) or commercial software such as Adobe Illustrator. Pajek files can also be used by other network analysis and visualization programs such as Gephi, VOSViewer, or UCINet. Note that one can zoom in on dense parts in the Draw screen of Pajek using the right mouse button for selecting specific areas in the map.

## 5. Conclusion

By providing basemaps for USPTO data in terms of the IPC and flexible tools for overlaying patent classifications on top of them both at the coarse-grained level of three digits and the fine-grained level of four digits, we have complemented our series of studies and tool development efforts to follow new development more systematically across databases. Grilliches (1994:14) already noted that the core problem in studying innovations has been the silos of data that are constructed for institutional reasons and then developed longitudinally over the years without sufficient cross-connections. For the study of innovations, one needs to map transversal translations from one context into another.

One common baseline among the different databases is provided by the institutional address that can be overlaid on Google Maps in order to show geographical diffusion (Leydesdorff & Rafols, 2011b). Additionally, one would like to move from one context to another in terms of the classifications and codifications that are proper to each database in cognitive terms. In the case of scholarly publishing, the prime unit of organization appears to have been the scientific journal. Using Web-of-Science or Scopus, citation relations can be accessed directly. In dedicated databases such as Medline for the bio-medical sciences, specific interfaces are needed to relate citation information to classifications (Leydesdorff & Opthof, in press).

The international patent classification IPC provides a means to organize patents intellectually, with a trade-off between technological refinement and user-friendliness. The IPC provides a baseline in all major patent systems such as WIPO, USPTO, and EPO. This paper has described the possibilities for using the fields in the USPTO databases for mapping and animation purposes. As an "instrumentality" (Price, 1984), the exploitation of this interface enables us to address central questions of patent analysis, such as those formulated in a study of technological competencies (Patel & Pavitt, 1997, at p. 141) concisely, as follows:

1. They [large firms] are typically *multi-field*, and becoming more so over time, with competencies ranging beyond their product range, in technical fields outside their 'distinctive core'.
2. They are *highly stable* and *differentiated*, with both the *technology profile* and the *directions* of localised search strongly influenced by firms' *principal products*.
3. The rate of search is influenced by both the firm's *principal products*, and the conditions in its *home country*. However, *considerable unexplained variance* suggests scope for managerial choice.

These conclusions identify a research program. The instruments provided here offer tools for addressing such a program of studies in quantitative terms (e.g., in terms of Rao-Stirling diversity) and for illustrating the results with animations of, for example, diffusion and diversification processes. Nowadays, the Internet enables us to upscale and use "big data" for performing these studies of science, technology, and innovation (Helbing & Balietti, 2011. In our opinion, the development of interfaces to access different databases ("big data") with flexibility, but with similar or equal search strings provides a strategy which may enable us to follow new developments in science and technology along trajectories and potentially developing into regimes (Leydesdorff *et al.*, in press).

In most previous studies using OECD data, for example, analysis remained at the aggregated level and static analysis consequently prevailed (e.g., Jaffe, 1989; Patel & Pavit, 1997). These new instruments enable us to study individual firms, nations, and new technologies in considerable detail and dynamically by following the available retrieval options and tracing the various classifications at USPTO. In a follow-up, we envisage studying. for example, the new classes of "nanotechnology" developed in the IPC and available as B82$ in USPTO,[4] and/or differently in the US Classification System as 977.[5] Such a study would allow one to follow the changing position of "nano-patents" in terms of IPC classifications over time.

**Acknowledgements**

We acknowledge support by the ESRC project 'Mapping the Dynamics of Emergent Technologies' (RES-360-25-0076). We are grateful to Nils Newman and Antoine Schoen for previous comments and communications.

---

[4] Since April 2009, class B82 in IPC replaces the previous class Y01 in ECLA.
[5] A search with "icl/B82$" provided a recall of 344 patents on September 25, 2012, whereas a search with "ccl/977$" provided a recall of 8,134 patents. Of these two sets 249 overlap (using an AND statement).